\documentclass[12pt]{iopart}

\usepackage{amssymb}
\usepackage{bm}
\usepackage{epsf}
\usepackage{graphicx}
\usepackage{cite}

\def\be{\begin{equation}}
\def\ee{\end{equation}}
\def\bea{\begin{eqnarray}}
\def\eea{\end{eqnarray}}

\def\ptl{\partial}

\begin{document}

\title [An inhomogeneous toy model of the quantum
gravity ] {An inhomogeneous toy model of the quantum gravity with
explicitly evolvable observables}

\author{ S.L. Cherkas\dag   \ and V.L. Kalashnikov\ddag
}

\address{\dag\
Institute for Nuclear Problems, Bobruiskaya 11, Minsk 220050,
Belarus}

\address{\ddag\ Technische Universit\"{a}t Wien, Gusshausstrasse 27/387, Vienna A-1040, Austria}

\begin{abstract}
An inhomogeneous (1+1)-dimensional model of the quantum gravity is
considered. It is found, that this model corresponds to a string
propagating against some curved background space. The quantization
scheme including the Wheeler-DeWitt equation and the ``particle on a
sphere'' type of the gauge condition is suggested. In the
quantization scheme considered, the ``problem of time'' is solved by
building of the quasi-Heisenberg operators acting in a space of
solutions of the Wheeler-DeWitt equation and the normalization of
the wave function corresponds to the Klein-Gordon type. To analyze
the physical consequences of the scheme, a (1+1)-dimensional
background space is considered for which a classical solution is
found and quantized.  The obtained estimations show the way to
solution of the cosmological constant problem, which consists in
compensation of the zero-point oscillations of the matter fields by
the quantum oscillations of the scale factor. Along with such a
compensation, a slow global evolution of a background corresponding
to an universe expansion exists.
\end{abstract}

\pacs{98.80.Qc, 11.10.-z, 11.25.Pm}

\section{Introduction}

The generally covariant canonical quantization of gravity has been
proposed in \cite{wheel,witt}. As a result, the well-known
Wheeler-DeWitt (WDW) equation has been obtained from the
Hamiltonian formulation of the general relativity theory (GR). The
main problem preventing a further development of the canonical
quantum gravity is an absence of explicit evolution that is a
consequence of the Hamiltonian constraint (e.g., see
\cite{CP,w,shest,hal,bar}).

It is well-known, that the string theory \cite{grin,brink} has the
Hamiltonian constraint, as well, but it does not face the
``problem of time''. The point is that either the equations of
motion are quantized (in the covariant approach) or the
time-dependent gauges are used (in the light-cone approach) in the
quantized string theory. From the other hand, the WDW equation is
not used in the string theory although this possibility was
discussed \cite{brink,mon}. A thought suggests itself, that the
quantization of the equations of motion can be suitable also in
the quantum gravity and, at the same time, the WDW equation can
make sense for some kinds of strings.

In our previous  works \cite{prep1,prep2}, we  have demonstrated
the quantization
scheme including both quantization of the equations of  motion
and the WDW equation for a minisuperspace model. The ideology of quantization  is quite
similar to that in the string theory, where some time-dependent
gauge is used, except the fact that the gauge is not imposed at all
times but only at the initial instant and thereafter it is permissible for operators
to evolve in accordance with the equations of motion. Another
distinctive feature of the method proposed is that the
evolvable operators (so-called, quasi-Heisenberg operators) act in a
space of the solutions of the WDW equation, which are normalized in the Klein-Gordon
style. Here we extend the quasi-Heisenberg quantization scheme to
the string in the curved background space that can be treated as an inhomogeneous (1+1)-dimensional toy model
of the quantum gravity.

\section{Model of a spatially inhomogeneous gravity}

We shall follow a heuristic method to obtain a spatially
inhomogeneous gravity model in two space-time dimensions.
The Lagrangian for a gravitation and a scalar field $\phi$ can be
written in the form
\begin{equation}
S=-\frac{M_p^2}{12}\int \mathcal G\sqrt{-g}\,d^4x+\frac{1}{2}\int
\biggl(\partial_\mu \phi\, g^{\mu\nu}\partial_\nu \phi -m^2\phi^2
\biggr)\sqrt{-g}\,d^4x, \label{sc}
\end{equation}
where $\mathcal G=g^{\alpha \beta } \left(\Gamma _{\alpha \nu
}^{\rho }
   \Gamma _{\beta \rho }^{\nu }-\Gamma _{\alpha \beta
   }^{\nu } \Gamma _{\nu \rho }^{\rho }\right)$ \cite{lan} and $M_p$ is the Planck mass,
    which is chosen as $M_p=\sqrt{\frac{3}{4\pi G}}$.

If one restricts a metric to the form
\[
ds^2=a^2(\tau,\bm r)(N^2(\tau,\bm r)d \tau^2-d\bm r^2),
\]
the resulting Lagrangian becomes:
\be
\eqalign{ \fl L=\frac{1}{2}\int
\biggl\{N\left(-\frac{M_p^2\,a^{\prime 2}}{N^2}+M_p^2(\bm \nabla
a)^2-\frac{2M_p^2}{3}\bm\nabla\cdot(a\bm\nabla a)
+a^2\biggl(\frac{\phi^{\prime
2}}{N^2}-(\bm\nabla\phi)^2\right)\nonumber\\~~~~~~~~~~~~~~~~~~~~~~~~
~~~~~~~~~~~~~~~~~-m^2a^4\phi^2\biggr)+\frac{2M_p^2}{3}\bm \nabla
\cdot(a\bm \nabla a \,N) \biggr\}d^3\bm r,}
\label{lagr}
\ee
where a prime denotes a derivative over the time.

The last term in Eq. (\ref{lagr}) can be omitted as a total
divergence. The variation with respect to $N$ gives the Hamiltonian constraint:
\begin{equation}
\fl \mathcal H=\frac{1}{2}\left(-\frac{M_p^2a^{\prime
2}}{N^2}-M_p^2(\bm \nabla
a)^2+\frac{2M_p^2}{3}\bm\nabla\cdot(a\bm\nabla a)
+a^2\frac{\phi^{\prime
2}}{N^2}+a^2(\bm\nabla\phi)^2+m^2a^4\phi^2\right)=0.
\end{equation}

Hereinafter, let us consider the conformal time gauge $N=1$. The term
$-\frac{2M_p^2}{3}\bm\nabla\cdot(a\bm\nabla a)$ in Eq.
(\ref{lagr}) does not affect the equations of motion, which have
the form
\bea
M_p^2a^{\prime\prime}-M_p^2\bm \nabla^2a+a \left(\phi^{\prime\,
2}-(\bm
\nabla \phi)^2\right)-2 a^3 m^2\phi^2=0,\\
\phi^{\prime\prime}+2\frac{a^\prime}{a}\phi^\prime-\frac{1}{a^2}\bm
\nabla\cdot(a^2\bm \nabla\phi)+a^2m^2\phi=0.
\eea

Then, the time evolution of the Hamiltonian constraint can be expressed as:
\be
\partial_\tau \mathcal H=\bm \nabla\cdot \mathcal {\bm P}
, \ee where the momentum constraint is $\mathcal {\bm
P}=-M_p^2a^\prime\bm \nabla a +a^2 \phi^\prime \bm \nabla \phi$. On
the contrary, the time derivative of the momentum constraint is not
expressed through $\mathcal H$ and $\mathcal {\bm P}$. Thus, this
system does not belong to the fist kind one \cite{dirac,han,git}
unlike GR. Nevertheless, if one considers an (1+1)-dimensional
analog of (\ref{lagr}) with a massless scalar field
\begin{equation}
L=\frac{1}{2}\int N\left(-\frac{M_p^2a^{\prime
2}}{N^2}+M_p^2\left(\frac{\partial a}{\partial \sigma}\right)^2
+a^2\left(\frac{{\phi^{\prime
2}}}{N^2}-\left(\frac{\partial\phi}{\partial
\sigma}\right)^2\right)\right)d\sigma,
\label{lag}
\end{equation}
($\sigma$ denotes  the only spatial variable), the Lagrangian
(\ref{lag}) proves to correspond to a string
propagating against some curved background space and, thus, it gives a
completely self-consistent system of constraints like that of GR.

\section{Connection with the string theory}

It is believed that the superstring theory shall include the GR. Nevertheless, one may try
to connect the string theory with the GR directly by means of
reduction of the GR Lagrangian on assumption that the space-time has some
symmetry \cite{mat}. In the present work, we shall demonstrate that the Lagrangian (\ref{lag}) with
a nonuniform scale factor describes a string against a curved background. Such a string model can be
interpreted as an inhomogeneous toy model for the quantum gravity.

Let us
write the standard form of action for a bosonic string \cite{grin} in a background
space
\begin{equation}\label{action}
S=\int d^2\xi\sqrt{-h}\,h^{\alpha\beta}(\xi)\partial_\alpha
X^\mu\partial_\beta X^\nu g_{\mu\nu} (X(\xi)),
\end{equation}
where $\xi=\{\tau,\sigma\}$. The metric tensor
$h_{\alpha\beta}(\xi)$ describes the intrinsic  geometry of
a (1+1)-dimensional manifold and $g_{\mu\nu}(X(\xi))$ describes the
geometry of a background space. Variation with respect to
$h_{\alpha\beta}(\xi)$ leads to the constraints \cite{grin}:
\begin{equation}
\frac{\delta S}{\delta h_{\alpha\beta}}\equiv T_{\alpha\beta}=
\partial_\alpha X^\mu\partial_\beta X^\nu
g_{\mu\nu}(X)-\frac{1}{2}h_{\alpha\beta}h^{\eta\kappa}\partial_\eta
X^\mu\partial_\kappa X^\nu \,g_{\mu\nu}(X)=0.
\label{9}
\end{equation}

\noindent If one takes $X^\mu=\{a,\phi\}$, the metric tensor  $h_{\mu\nu}$
in the form of

\[h= \left(
\begin{array}{cc}
-N^2+N_1^2&  N_1
\\
N_1 &1
\end{array}
\right), \]  and the metric tensor $g_{\mu\nu}(X)$ of a background space
 as
\be g= \left(
\begin{array}{cc}
M_p^2  &0
\\
0 &-a^2
\end{array}
\right),
\label{bmetr}
 \ee
  it results in the
Lagrangian for the (1+1)-dimensional model
\bea
\fl L=\int\Biggl(\frac{M_p^2}{2}\partial_\sigma a^2 \left(
{N}-\frac{{N_1}^2}{ {N}}\right)+M_p^2\frac{\partial_\sigma a\,
   a^\prime {N_1}}{{N}}-M_p^2\frac{a^{\prime 2}}{2 {N}}\nonumber\\~~~~~~~~~~~~~~-\frac{a^2
   {N_1}\partial_\sigma
   \phi\,
    \phi ^{\prime}}{{N}}+\frac{1}{2}a^2\,\partial_\sigma\phi^2 \left(\frac{ {N_1}^2}{
   {N}}- {N}\right)+\frac{a^2 \phi ^{\prime 2}}{2
   {N}}\Biggr)d\sigma.
   \label{ll}
\eea The substitutions of $N_1=0$ and $N=1$ into (\ref{ll}) reduces
it to (\ref{lag}). Thus, the constraints (\ref{9}) become the
Hamiltonian and momentum constraints of the model considered.

In a more general case, one may consider a number of scalar fields
besides the scale factor $a$, i.e
$X=\{a,\phi_1,\phi_2,\dots\phi_{N}\}$ and take the background
metric tensor
\be
g_{\mu\nu}(X)=\mbox{diag}\{M_p^2,-a^2,-a^2\dots\}.
\label{bmetr1}
\ee

Let us denote $\alpha=\ln a$ and rewrite the Lagrangian
(\ref{lag}) in the terms of $\alpha$ and a set of scalar fields :

\be \fl L=\frac{1}{2}\int e^{2\alpha}\left(-M_p^2\alpha^{\prime
2}+\bm\phi^{ \prime
2}+M_p^2(\partial_\sigma\alpha)^2-(\partial_\sigma\bm
\phi)^2\right)d \sigma=0, \ee where $\bm \phi=\{\phi_1,\phi_2
\dots\phi_N \}$. The relevant Hamiltonian and momentum constraints
(\ref{9}), written in the terms of momentums $\bm
\pi(\sigma)\equiv\frac{\delta L}{\delta {\bm \phi^\prime(\sigma)
}}=e^{2\alpha}\bm \phi^\prime$ and
$p_\alpha(\sigma)\equiv-\frac{\delta L}{\delta \alpha
^\prime(\sigma)}=M_p^2\,e^{2\alpha}\alpha^\prime$ take the form of
\begin{eqnarray}
\mathcal{H}(\sigma)=\frac{1}{2}e^{-2\alpha}\left(-p_\alpha^{2}/M_p^2+\bm\pi^{
2}\right)+e^{2\alpha}\left(-M_p^2(\partial_\sigma\alpha)^2+(\partial_\sigma\bm \phi)^2\right)=0,\label{sv0}\\
\mathcal{P}(\sigma)=-p_\alpha\partial_\sigma\alpha+\bm
\pi\partial_\sigma\bm \phi=0.
\label{sv}
\end{eqnarray}
The constraints algebra demonstrates that there are no new
constraints:
\begin{eqnarray*}
\{\mathcal H(\sigma),\mathcal H(\sigma^\prime)\}=(\mathcal
P(\sigma)+\mathcal
P(\sigma^\prime))\delta^\prime(\sigma-\sigma^\prime), \\
\{\mathcal P(\sigma),\mathcal P(\sigma^\prime)\}=(\mathcal
P(\sigma)+\mathcal
P(\sigma^\prime))\delta^\prime(\sigma-\sigma^\prime),\\
\{\mathcal P(\sigma),\mathcal H(\sigma^\prime)\}=(\mathcal
H(\sigma)+\mathcal
H(\sigma^\prime))\delta^\prime(\sigma-\sigma^\prime),
\end{eqnarray*}
where $\delta^\prime(\sigma-\sigma^\prime)$ is the derivative of the
Dirac delta-function, and the Poisson brackets are implied: \be
\eqalign{ \fl\{A(\sigma),B(\sigma^\prime)\}=\int\Biggl(-\frac{\delta
A(\sigma)}{\delta p_\alpha(\xi)}\frac{\delta
B(\sigma^\prime)}{\delta \alpha(\xi)}+\frac{\delta A(\sigma)}{\delta
\alpha(\xi)}\frac{\delta B(\sigma^\prime)}{\delta
p_\alpha(\xi)}\\+\sum_{j}^N\frac{\delta A(\sigma)}{\delta
\pi_j(\xi)}\frac{\delta B(\sigma^\prime)}{\delta
\phi_j(\xi)}-\frac{\delta A(\sigma)}{\delta \phi_j(\xi)}\frac{\delta
B(\sigma^\prime)}{\delta \pi_j(\xi)}\Biggr)d\xi}\label{pois} \ee .

The next step is the quantization of the model presented. It is
known, that the string theory demands $X^{\mu}$ to have the critical
dimension, which is of 26 for a bosonic string. Nevertheless, even
for the critical dimension, one cannot expect that the string theory
gives a satisfactory quantization of the model against a curved
background. The point is that the string theory provides a
consistent quantization only if the Ricci tensor $R_{\mu\nu}(X)$ of
the background metric $g_{\mu\nu}(X)$ equals to zero
\cite{grin,mukhi}. For $g_{\mu\nu}$ given by (\ref{bmetr1}), it is
nonzero if the dimension of $X^{\mu}$ is higher than two (the scalar
curvature is constant in the case considered). However, it was
stated \cite{grin, mukhi} that an inclusion of some other fields
(e.g., dilaton) leads to the standard Einstein equations
$R_{\mu\nu}(X) \neq 0$ for the background metric and it can be
considered as a new principle of derivation of the Einstein
equations \cite{mukhi}.

It should be noted, that the string representation arises in our
paper in a quite different context than the ordinary one (in
particular, as used in \cite{mat}). In our model the background
space is an isospin-like space determined by the number of scalar
fields (matter degrees of freedom, whereas the physical space,
where the observers and detectors could be situated) is a $(\tau,
\sigma)-$ space. On the contrary, the ordinary string theory
considers a background space as ``physical'' one, whereas a
$(\tau, \sigma)-$space corresponds to ``internal degrees of
freedom''. Thus, the quantization resulting in the dimension
constraints for a background space (i.e., for a number of matter
fields in our case) can be hardly acceptable in the model
considered. In the next section, we shall present the quantization
scheme for $R_{\mu\nu}(X)\ne 0$ and a background space without any
dimension constraints.

\section{Quantization}

\subsection{Relativistic ``fluid'' on a sphere}

The minisuperspace model presented in \cite{prep1,prep2}
demonstrates that the quasi-Heisenberg quantization consists in the
following: one has to set the initial values for the operators,
which evolve in accordance with the operator equations of motion.
The question is to construct the representation for these operators
and the space where they act. In our approach this problem is solved
by consideration of the Hamiltonian constraint as the WDW equation
and choice of the plane $a=0$ for normalization of a wave function
in the Klein-Gordon style. In the general case, the momentum
constraint is also should be taken into account. This requires to
impose one additional gauge condition. The light cone gauge
\cite{grin,goddard}
 is
widely used in the string theory. Since it contains necessarily a
nonzero mean momentum, it is hardly acceptable for our isotropic
cosmological-like model. Thus, it is necessary to propose some new
gauge. The hint arises from the form of the momentum constraint,
which reminds that in the theory of a particle moving on a sphere
\cite{klein,Neto,Scar,Golov}.

At first, let us consider a simpler problem described by the
Hamiltonian \be H=\frac{1}{2}(-p_0^2+\bm p^2)=0, \label{gp} \ee
which is similar to that of a massless relativistic particle.

The analog of the momentum constraint can be writing as \be
 P=-x_0p_0+\bm p\,\bm r=0,
\ee
where $\bm p=\{p_1,p_2\dots p_N\}$, $\bm r=\{x_1,x_2\dots x_N\}$.

The constraint algebra is
\begin{equation}
\{H,P\}=-\frac{\partial H}{\ptl p_0}\frac{\ptl P}{\ptl
x_0}+\frac{\ptl H}{\ptl x_0} \frac{\ptl P}{\ptl p_0}+\frac{\ptl
H}{\ptl \bm p}\frac{\ptl P}{\ptl \bm r}-\frac{\ptl H}{\ptl \bm r
}\frac{\ptl P}{\ptl \bm p}=2 H.
\end{equation}

 Let us
take the gauge condition in the form of
\begin{equation}
{\mathcal A}=-x_0^2+\bm r^2-R^2=0,
\end{equation}
where $R$ is some constant.

\begin{figure}[h]
\vspace{0. cm} \hspace{.5 cm}
 \includegraphics[width=7.5 cm]{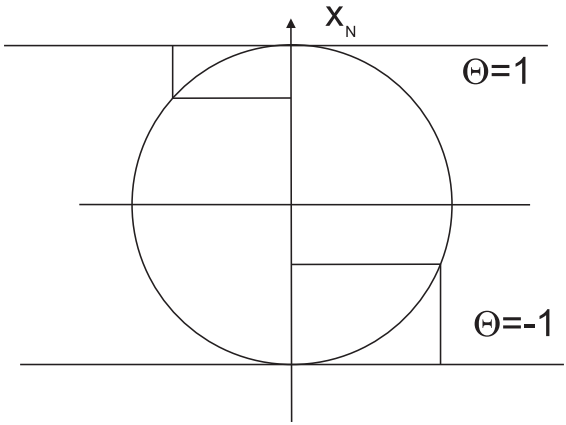}
 \vspace{. cm}
 \caption{Vertical plane projections of the points on a sphere.} \label{ris}
 \end{figure}

The first step is to formulate the WDW equation. The momentum
constraint and the gauge condition allow excluding one of the
degrees of freedom, for instance, the coordinate $x_N$ and the
momentum $p_N$. Let's denote $\bm \Pi=\{p_1,p_2\dots p_{N-1}\}$ and
$\bm {\mathcal X}=\{x_1,x_2\dots x_{N-1}\}$. Then \bea
x_N=\Theta\,\sqrt{R^2+x_0^2-\bm {\mathcal X}^2}\label{25},\\
p_N=\Theta\,(x_0 p_0-\bm {\mathcal X}\,\bm
\Pi)/\sqrt{R^2+x_0^2-\bm {\mathcal X}^2}, \label{26} \eea where
$\Theta$ takes the values $\pm 1$. The coordinates $\bm {\mathcal
X}$ and $\Theta$ correspond to the vertical plane projection of a
sphere shown in Fig. \ref{ris}. The value $\Theta=1$ corresponds
to the upper semisphere and $\Theta=-1$ corresponds to the lower
one.

Substitution of the above quantities into (\ref{gp}) results in

\begin{equation}
-p_0^2+\bm \Pi^2+\frac{(p_0 x_0-\bm \Pi\bm {\mathcal X
})^2}{R^2+x_0^2-\bm {\mathcal X}^2}=0.
\label{ww}
\end{equation}

Further, we shall take an interest only in the region of
$x_0\rightarrow 0$ and $R\rightarrow \infty$, where the asymptotic
of Eq. (\ref{ww}) takes the form of

\begin{equation}
-p_0^2+\bm \Pi^2=0.
\label{ww1}
\end{equation}
Obtaining of the WDW equation consists in the postulation of
\[
p_0=i\frac{\partial}{\partial x_0},~~~~~ \bm
\Pi=-i\frac{\partial}{\partial \bm {\mathcal X}},
\]
that results in
\begin{equation}
\left(\frac{\partial^2}{\partial x_0^2
}-\frac{\partial^2}{\partial \bm {\mathcal X}^2
}\right)\Psi(x_0,\Theta,\bm {\mathcal X} )=0.
\end{equation}
The wave  packet solution \cite{prep1,prep2} is\footnote{In the
general context, an analog of the plane waves on a sphere
\cite{sherman,vol,alonso,red} can be used.}

\begin{equation}
\Psi(x_0,\bm {\mathcal X} ,\Theta)=\int c(\bm k,\Theta)e^{-i |\bm
k|x_0+i\bm k \bm {\mathcal X}}d^{N-1} \bm k ,
\end{equation}
where $c(\bm k,\Theta)=c_0(\bm k)+\Theta\, c_1(\bm k)$.

A mean value of an arbitrary operator $A$ is given by
\cite{prep1,prep2,mostf}

\begin{eqnarray}
\fl <\Psi|\hat A(x_0,\Theta,\bm {\mathcal X} ,\hat {\bm
\Pi})|\Psi>= \frac{i}{2}\sum_{\Theta=\pm 1}\int \Biggl(
\Psi^*(x_0,\Theta,\bm{\mathcal X}) {\hat D^{1/4}}\hat A\,{\hat
D^{-1/4}}\frac{\partial}{\partial x_0 }\Psi(x_0,\Theta,\bm
{\mathcal X} )\nonumber\\-\left(\frac{\partial}{\partial x_0
}\Psi^*(x_0,\Theta,\bm {\mathcal X})\right){\hat D^{-1/4}}\hat A
\,{\hat D^{1/4}} \,\Psi(x_0,\Theta,\bm {\mathcal X})\Biggr)
d^{N-1} \bm{\mathcal X} \biggr|_{\,x_0\rightarrow 0},
\label{meanv1}
\end{eqnarray}
where  $\hat D$ is the operator $ \hat
D=-\frac{\partial^2}{\partial \bm {\mathcal X}^2}$.

In the momentum representation of the wave function
\[
\Psi(x_0,\bm {\mathcal X} ,\Theta)=\int c(\bm k,\Theta )e^{-i|\bm
k|x_0+i \bm k \bm {\mathcal X}}d^{N-1}\bm k,
\]
the mean value of an operator  is given as
\begin{equation}
\fl < \hat A
>=\sum_{\Theta=\pm 1}\int c^*(\bm k,\Theta )e^{i|\bm k|x_0} \hat
A(x_0,\Theta,i\frac{\partial}{\partial \bm k} , {\bm k})\,c(\bm
k,\Theta)e^{-i|\bm k|x_0}\,d^{N-1}\bm k \biggr|_{\,x_0\rightarrow
0}.
\label{meanv2}
\end{equation}

The equations of motion follow from Eq. (\ref{gp}) via the
Hamilton-Jacobi equations $\frac{d x_i}{d\tau}=\{H,x_i\}$, $\frac{d
p_i}{d\tau}=\{H,p_i\}$ and take the form of \be \frac{d
x_i}{d\tau}=p_i,~~~ \frac{d x_0}{d\tau}=p_0,~~~
\frac{dp_i}{d\tau}=0,~~~ \frac{dp_0}{d\tau}=0. \label{xmov} \ee We
shall consider them as the operator equations. For more complicated
systems, some operator ordering could be needed.

Now one has to assign the initial values for the quasi-Heisenberg
operators. In our model, they are build according to the Dirac rule
for the full set of constraints and gauge conditions but only at the
initial moment of the proper time $\tau$. The full set of
constraints at $\tau=0$ is $F=\{H,P,{\mathcal A},{\mathcal B}\}$,
where the gauge condition ${\mathcal B}=x_0$ suggests $x_0(0)=0$.
This gauge condition is taken to adjust the commutation rules for
the initial values of quasi-Heisenberg operators to the hyperplane
$x_0=0$, where a wave function is normalized in the Klein-Gordon
style in accordance with Eqs. (\ref{meanv1}), (\ref{meanv2}).

The Dirac matrix consists of the Poisson brackets of constraints and
the gauge conditions

\be
\fl M_{\alpha\beta}=\{F_\alpha,F_\beta\}=-\frac{\partial
F_\alpha}{\partial p_0}\frac{\partial F_\beta}{\partial
x_0}+\frac{\partial F_\alpha}{\partial x_0}\frac{\partial
F_\beta}{\partial p_0}+\sum_{j}^N\frac{\partial F_\alpha}{\partial
p_j}\frac{\partial F_\beta}{\partial x_j}-\frac{\partial
F_\alpha}{\partial x_j}\frac{\partial F_\beta}{\partial p_j},
\ee

\noindent that leads to
\be
M=\left(
\begin{array}{cccc}
0&0&0&p_0\\
0&0&2R^2&0\\
0&-2R^2&0&0\\
-p_0&0&0&0
\end{array}
\right).
\ee

The inverse matrix $M^{-1}$ is evident and we can calculate the
Dirac brackets \cite{dirac,han,git}:
\be
\{p_i,x_j\}_D=\{p_i,x_j\}-\sum_{\alpha,\beta}\{p_i,F_\alpha\}M^{-1}_{\alpha\beta}\{F_\beta,x_j\}.
\label{pij}
\ee
The corresponding commutation relations at $\tau=0$ can be  obtained
by multiplication of the right hand side of Eq. (\ref{pij}) by
$(-i)$ that results in

\be
\eqalign{
 \left[\hat p_l(0),\hat
x_j(0)\right]=-i \left(\delta_{lj}-\frac{\hat x_l(0)\hat
x_j(0)}{R^2}\right),\cr \left[\hat x_i(0),\hat
x_j(0)\right]=0,~~~~ \left[\hat p_l(0),\hat
p_j(0)\right]=\frac{-i}{R^2}S\left(\hat p_l(0) \hat x_j(0)-\hat
x_l(0) \hat p_j(0)\right),\cr \left[\hat p_0^2(0),\hat
x_j(0)\right]=-i\frac{2}{R^2}S\left(\hat p_j(0)\sum_{m\ne j}^N\hat
x_m^2(0) -\hat x_j(0)\sum_{m\ne j}^N\hat p_m(0) \hat
x_m(0)\right),\cr \left[\hat p_0^2(0),\hat
p_j(0)\right]=-i\frac{2}{R^2}S\left(\hat x_j(0)\sum_{m\ne j}^N\hat
p_m^2(0) -\hat p_j(0)\sum_{m\ne j}^N\hat p_m(0) \hat
x_m(0)\right),}\label{comr}
\ee
where symbol $S$ denotes symmetrization of the noncommuting
operators, i.e. $S(\hat A \hat B)=\frac{1}{2}(\hat A\hat B+\hat
B\hat A )$ or $S(\hat A\hat B\hat C)=\frac{1}{6}(\hat A\hat B\hat
C+\hat B\hat A\hat C+\hat A\hat C\hat B+\dots)$. Certainly, the
quantity $\hat x_0(0)=0$ is $c$-number and commutes with all
operators.

 Realization
of the commutation relations is represented as
\be
\fl\eqalign{\hat x_j(0)=x_j,~~~~~
 \hat p_j(0)=-i\left(\frac{\partial
}{\ptl x_j}-\frac{1}{R^2}S\left(x_j\left(\bm{\mathcal
X}\frac{\partial }{\ptl \bm{\mathcal X}}\right)\right)\right),~~~~
j\in\{1,N-1\},\\
\hat x_N(0)=\Theta\sqrt{R^2-\bm{\mathcal X}^2},~~~~ \hat
p_N(0)=i\,S\left(\frac{\Theta\sqrt{R^2-\bm{\mathcal
X}^2}}{R^2}\left(\bm{\mathcal X}\frac{\partial }{\ptl \bm{\mathcal
X}}\right)\right),\\ \hat p_0(0)=\sqrt{\sum_{j=1}^{N} \hat p_j^2(0)}
.}\label{rlz} \ee

Let us emphasize, that all commutation relations and constraints are
satisfied as the operator equalities (i.e. strongly) at the initial
time if the symmetrized quantum version of $\hat {\mathcal
P}(0)=S\left( \sum_{j=1}^N \hat p_j(0)\hat x_j(0)\right)\ $ is
implied for the ${\mathcal P}$ constraint.

Now one can solve the equations of motion (\ref{xmov}) with the
above initial conditions (\ref{rlz}) and calculate the mean values of
operators according to (\ref{meanv1}), (\ref{meanv2}).

It is essential, that Eqs. (\ref{rlz}) can be simplified in the vicinity of $R\rightarrow \infty$:

\be
\eqalign{\hat x_j(0)=x_j,~~~~
 \hat p_j(0)=-i\frac{\partial
}{\ptl x_j},~~~~ j\in\{1,N-1\},\\
\\
\hat x_N=\Theta\,R,~~~~ \hat
p_N(0)=i\,\frac{\Theta}{R}S\left(\bm{\mathcal X}\frac{\partial
}{\ptl \bm{\mathcal X}}\right).}\label{rlz1}
\ee

\subsection{Quantization scheme for the string in a (1+N)-dimensional space}

We intend to quantize the string in a curved background space by
close analogy with the quantization of the ``fluid'' on a sphere.
Let's take the gauge condition in the form of

\begin{equation}
{\mathcal
A}=e^{2\alpha}\left(-M_p^2(\partial_\sigma\alpha)^2+(\partial_\sigma\bm
\phi)^2\right)-\Lambda^2=0,
\label{usl1}
\end{equation}
where $\Lambda$ is some constant. In fact, we equal the part of
Hamiltonian density to $\Lambda^2$ at the initial time.

Let's $\bm \Pi=\{\pi_1,\pi_2\dots\pi_{N-1}\}$, $\bm
\Phi=\{\phi_1,\phi_2\dots\phi_{N-1}\}$, then $\phi_N$ and $\pi_N$ are expressed through
the momentum constraint (\ref{sv}) and the gauge condition (\ref{usl1})
 as
\begin{eqnarray}
\partial_\sigma\phi_N(\sigma)=\Theta(\sigma)\sqrt{\Lambda^2e^{-2\alpha}+M_p^2(\partial_\sigma\alpha)^2-(\partial_\sigma \bm
\Phi)^2},\label{1}
\\
\pi_N(\sigma)=\Theta(\sigma)\frac{p_\alpha\partial_\sigma
\alpha-\bm \Pi\partial_\sigma\bm
\Phi}{\sqrt{\Lambda^2e^{-2\alpha}+M_p^2(\partial_\sigma\alpha)^2-(\partial_\sigma
\bm \Phi)^2}},\label{2}
\end{eqnarray}
where $\Theta(\sigma)$ is the piecewise function taking the values
$\pm 1$ such that $\Theta^2(\sigma)=1$.

Substitution of Eqs. (\ref{1}),(\ref{2}) to (\ref{sv0}) leads  to
\begin{equation}\label{39}
e^{-2\alpha}\left(-p_\alpha^2/M_p^2+\bm
\Pi^2+\frac{(p_\alpha\partial_\sigma\alpha-\bm
\Pi\partial_\sigma\bm
\Phi)^2}{\Lambda^2e^{-2\alpha}+M_p^2(\partial_\sigma\alpha)^2-(\partial_\sigma
\bm \Phi)^2}\right)+\Lambda^2=0.
\end{equation}

The asymptotic of Eq. (\ref{39}) in the vicinity of
$\alpha(\sigma)=\alpha_0\rightarrow-\infty$ is
\begin{equation}
e^{-2\alpha}\left(-p_\alpha^2/M_p^2+\bm \Pi^2\right)=0.
\end{equation}
Hence, the quantization leads to the WDW equation
\begin{equation}
\left(\frac{1}{M_p^2}\frac{\delta^2}{\delta
\alpha(\sigma)}-\left(\frac{\delta^2}{\delta^2 \bm
\Phi(\sigma)}\right)\right)\Psi[\alpha,\Theta,\bm
\Phi]=0,\label{wi}
\end{equation}
where $\frac{\delta}{\delta f(\sigma)}$ denotes the functional
derivative.

The solution of Eq. (\ref{wi}) is  the functional, which can be
written as
\begin{equation}
\Psi[\alpha,\Theta,\bm \Phi]=\int C[\bm {\mathcal
K},\Theta]\,e^{\int\left(-i M_p|\bm {\mathcal
K}(\sigma)|\alpha(\sigma)+i\bm {\mathcal K}(\sigma)\bm
\Phi(\sigma)\right)d\sigma}\,\mathcal D \bm {\mathcal K},
\end{equation}
where  $\int\dots \mathcal D \bm {\mathcal K}$ denotes a
functional integral.

A mean value of an arbitrary operator can be evaluated as
\cite{prep1,prep2,mostf}
\be
\eqalign{ \fl <\Psi|\hat A[\alpha,-i\frac{\delta}{\delta \bm
\Phi},\bm \Phi]|\Psi>=i \int \biggl( \Psi^*[\alpha,\Theta,\bm
\Phi] {\hat D^{1/4}}\hat A\,{\hat D^{-1/4}}\frac{\delta}{\delta
\alpha(\sigma) }\Psi[\alpha,\Theta,\bm
\Phi]\\-\left(\frac{\delta}{\delta \alpha(\sigma)
}\Psi^*[\alpha,\Theta,\bm \Phi]\right){\hat D^{-1/4}}\hat A\,{\hat
D^{1/4}}\,\Psi[\alpha,\Theta,\bm \Phi]\biggr)\mathcal D \bm
\Phi\,\mathcal D
\Theta\biggr|_{\,\alpha(\sigma)=\alpha_0\rightarrow -\infty
},\label{mean3}
 }
\ee
where $\hat D(\sigma)=\frac{\delta^2}{\delta \bm \Phi^2(\sigma)}$.

The classical equations of motion obtained with the help of the
Poisson brackets (\ref{pois})
  are written as
\be
\eqalign{ \bm \phi^{\prime\prime}-\partial_{\sigma\sigma}\bm
\phi+2\alpha^\prime\bm \phi^\prime
-2\partial_{\sigma}\alpha\partial_{\sigma}\bm \phi=0,\nonumber\\
M_p^2\alpha^{\prime\prime}-M_p^2\partial_{\sigma\sigma}\alpha+M_p^2\alpha^{\prime
2}-M_p^2(\partial_\sigma\alpha)^2+\bm \phi^{\prime
2}-(\partial_\sigma\bm \phi)^2=0.}
\label{eqns1}
\ee

Again, we shall consider these equations as the equations for the
quasi-Heisenberg operators.  One has to write the initial conditions
for these equations and to choose the operator ordering in
(\ref{eqns1}).

The additional gauge condition corresponding to the choice of the
hyperplane $\mathcal \alpha(\sigma)=\alpha_0\rightarrow\infty$ in
Eq. (\ref{mean3}) is
\be \mathcal B=\alpha(0,\sigma)-\alpha_0=0, \label{usl2} \ee where
$\alpha_0$ is $c$-number which will be tended to minus infinity.
Thus, there are two constraints $\mathcal H$, $\mathcal P$ given
by Eqs. (\ref{sv0}), (\ref{sv}) and two gauge conditions $\mathcal
A$ , $\mathcal B$ given by Eqs. (\ref{usl1}), (\ref{usl2}), which
are imposed at $\tau=0$. This allows calculating the Dirac
brackets and obtaining the commutation relations at the initial
moment of time. However, since it is necessary to know a
representation of operators only in the vicinity of
$\alpha_0\rightarrow -\infty$ where the radius of a sphere is
$\Lambda e^{-\alpha_0}$ and , thereby, is infinitely large, one
can write by analogy with (\ref{rlz1}) \be \fl\eqalign{ \hat
\phi_j(0,\sigma)={\phi}_j(\sigma),~~~~\hat
\pi_j(0,\sigma)=-i\frac{\delta }{\delta \phi_j},~~~~
j\in\{1,N-1\},\\
\hat \phi_N(0,\sigma)=\Delta(\sigma)\,\Lambda\, e^{-\alpha_0},~~~~
\hat \pi_N(0,\sigma)=i\,\frac{\Theta(\sigma)}{\Lambda \,
e^{-\alpha_0}}S\left(\bm{ \Phi}(\sigma)\frac{\delta }{\delta \bm{
\Phi}(\sigma)}\right),\\
\hat \alpha(0,\sigma)=\alpha_0,~~~~~ \hat
p_\alpha(0,\sigma)=M_p\sqrt{\sum_{j=1}^{N} \hat
\pi_j^2(0,\sigma)+\exp(2\alpha_0)\Lambda^2},} \label{rlz2} \ee
where
$\Delta(\sigma)$ is the piece linear function such that
$\Theta(\sigma)=\partial_\sigma \Delta(\sigma)$. The initial values
of time derivatives of the scale factor logarithm and scalar field
are expressed through the momentums given by (\ref{rlz2}) as $\hat
\alpha^\prime(0,\sigma)=e^{-2\alpha_0}\hat p_\alpha(0,\sigma)$ and
$\hat \phi_j^\prime(0,\sigma)=e^{-2\alpha_0}\hat \pi_j(0,\sigma)$.

 The last step is to choose the operator ordering in the
equations of motion (\ref{eqns1}). This problem is closely related
to the constraint evolution. Heretofore, it was guessed that the
constraints are satisfied only at the initial moment of time. One
can ask whether the constraints will be satisfied during an
evolution. Below we prove that if the symmetric (Weyl) ordering is
chosen in the expressions for the constraints then there exists the
operator ordering for the equations of motion, which preserves the
constraints during an evolution.

One needs to extend the symmetrization operation $S$ used in
(\ref{comr}) to a more general non-polynomial case \cite{masl,kar}.
 Let's there is an arbitrary function $f(x_1, x_2, \cdots x_n)$ of
 $n-$
 variables.  Then one can define the formal Fourier transform
\bea \fl \tilde
f(\zeta_1,\zeta_2\dots\zeta_n)=\frac{1}{(2\pi)^n}\int
f(x_1,x_2,\dots x_n)e^{-i\left(x_1\zeta_1+x_2\zeta_2\dots x_n\zeta_n
\right)}d x_1\dots d x_n. \eea The symmetrized function of the
noncommuting operators $\hat A_1, \dots \hat A_n $ is defined as \be
\fl S(f(\hat A_1,\hat A_2\dots \hat A_n))=\int \tilde
f(\zeta_1,\zeta_2,\dots \zeta_n)e^{i(\hat A_1\zeta_1+\hat
A_2\zeta_2\dots \hat A_n\zeta_n )}d \zeta_1\dots d \zeta_n. \ee The
remarkable formula for the differentiation of the symmetrized
function exists \cite{kar} \be \fl\frac{d}{dt} S(f(\hat A_1(t),\hat
A_2(t)\dots \hat A_n(t)))=S\left(\sum_j^n \frac{d \hat A_j}{d
t}\,{\partial_j} f(\hat A_1(t),\hat A_2(t)\dots \hat A_n(t))\right),
\label{differ} \ee where $\partial_j f(x_1,x_2\dots x_n)$ denotes a
partial derivative of the function $f$ over the $j-$argument. Let us
define the operator constraints as \bea \hat {\mathcal
H}=S\left(e^{2\hat\alpha}\left(-M_p^2\hat \alpha^{\prime
2}-M_p^2(\partial_\sigma\hat\alpha)^2+\hat{\bm\phi}^{\prime
2}+(\partial_\sigma\hat{\bm\phi})^2\right)\right),\nonumber\\
\hat {\mathcal
P}=S\left(e^{2\hat\alpha}\left(-M_p^2\hat\alpha^\prime\,\partial_\sigma\hat\alpha
+\hat{\bm\phi}^\prime\,\partial_\sigma\hat{\bm\phi}\right)\right).
\eea Then, let's calculate the quantities $\partial_\sigma\hat
{\mathcal H}$,
 $\partial_\sigma\hat {\mathcal  P}$, $\partial_\tau\hat {\mathcal
 H}$ and
 $\partial_\tau\hat {\mathcal  P}$ with the help of Eq. (\ref{differ})
 and consider the following equations
\be
\eqalign{
\partial_\tau\hat{\mathcal H}-\partial_\sigma\hat{\mathcal P}=0,\\
\partial_\tau\hat{\mathcal P}-\partial_\sigma\hat{\mathcal H}=0,}
\label{prop} \ee as two operator equations of motion, namely: \be
\eqalign{ \fl S\biggl(-\hat
\alpha^{\prime}e^{2\hat\alpha}\left(M_p^2\hat\alpha^{\prime\prime}
-M_p^2\partial_{\sigma\sigma}\hat\alpha+M_p^2\hat\alpha^{\prime
2}-M_p^2(\partial_\sigma\hat\alpha)^2+ \hat{\bm\phi}^{\prime
2}-(\partial_\sigma\hat{\bm \phi})^2)\right)\nonumber\\+
e^{2\hat\alpha}\hat{\bm \phi}^{\prime}\left(\hat{\bm
\phi}^{\prime\prime}-\partial_{\sigma\sigma}\hat{\bm
\phi}+2\hat\alpha^\prime\hat{\bm \phi}^\prime
-2\partial_{\sigma}\hat\alpha\partial_{\sigma}\hat{\bm
\phi}\right)\biggr)=0,\\
\fl S\biggl(-\ptl_\sigma\hat\alpha
\,e^{2\hat\alpha}\left(M_p^2\hat\alpha^{\prime\prime}
-M_p^2\partial_{\sigma\sigma}\hat\alpha+M_p^2\hat\alpha^{\prime
2}-M_p^2(\partial_\sigma\hat\alpha)^2+\hat{\bm \phi}^{\prime
2}-(\partial_\sigma\hat{\bm \phi})^2)\right)\nonumber\\+
e^{2\hat\alpha}\ptl_\sigma\hat{\bm \phi}\left({\hat{\bm
\phi}^{\prime\prime}-\partial_{\sigma\sigma}\hat{\bm
\phi}}+2\hat\alpha^\prime\hat{\bm \phi}^\prime
-2\partial_{\sigma}\hat\alpha\,\partial_{\sigma}\hat{\bm
\phi}\right)\biggr)=0.} \label{eqNN} \ee The remaining operator
equations of motion can be written as \be
\hat\phi_j^{\prime\prime}-\partial_{\sigma\sigma}
\hat\phi_j+2S(\hat\alpha^\prime\hat \phi_j^\prime)
-2S(\partial_{\sigma}\hat\alpha\,\partial_{\sigma} \hat\phi_j)=0,
~~~~j\in\{1,N-1\}. \label{NN} \ee

Eqs. ({\ref{eqNN}}) result in the constraint evolution in accordance
with Eqs. (\ref{prop}). The lasts have a trivial solution if and
only if initially $\hat \mathcal H=0$, $\hat \mathcal P=0$ . Thus if
the constraints were satisfied initially, they will be satisfied
during an evolution.  It should be emphasized, that Eqs.
({\ref{eqNN}}), (\ref{NN}) are completely equivalent to Eqs.
(\ref{eqns1}) for the commuting quantities (i.e., in the classics).

Thus, we have formulated an exact quantization scheme consisting
of the equations of motion (\ref{eqNN}), (\ref{NN}), the initial
conditions for the operators (\ref{rlz2}) and the formula
(\ref{mean3}) for calculation of the mean values. A distinctive
feature of this scheme is quantization of the scale factor
entangled with other degrees of freedom through the initial
condition for the momentum $\hat p_\alpha(\tau,\sigma)$.

It would be interesting to consider the physical consequences of
such a scale factor quantization. The main problem is to  solve the
operator equations of motion. For the minisuperspace model, this
problem has been solved numerically \cite{prep1,prep2} by using the
Weys symbols of operators. The problem becomes simpler if a system
possesses an analytical solution. For a string against a background,
the closed form of solution can be found only if a background space
is $(1+1)$-dimensional. But in this case, there exist no independent
degrees of freedom and the solutions are piecewise function like
$\Theta(\sigma)$ and $\Delta(\sigma)$ \cite{6s}.

\subsection{Some estimations for the string in a (1+1)-dimensional background space}

To estimate the physical consequences of the model considered,
let's consider, as before, the string in a (1+1)-dimensional
background space. In order to preserve at least one real degree of
freedom in this case, it is necessary to weaken the constraints.

The analytical solutions of the equations of motion (\ref{eqns1})  can
be written as
\be
\eqalign{ \fl \phi(\tau,\sigma)=\frac{1}{2}M_p
\biggl\{\ln\biggl(\frac{1}{2}
e^{\frac{1}{M_p}\varphi(\sigma-\tau)}+\frac{1}{2}
e^{\frac{1}{M_p}\varphi(\sigma+\tau)}+\frac{e^{-2\alpha_0}}{2M_p}\int_{\sigma-\tau}^{\sigma+\tau}(\pi(\xi)+|\pi(\xi)|)
e^{\frac{1}{M_p}\varphi(\xi)}d\xi\biggr)\nonumber\\\fl-\ln\biggl(\frac{1}{2}
e^{-\frac{1}{M_p}\varphi(\sigma-\tau)}+\frac{1}{2}
e^{-\frac{1}{M_p}\varphi(\sigma+\tau)}+\frac{e^{-2\alpha_0}}{2M_p}\int_{\sigma-\tau}^{\sigma+\tau}(|\pi(\xi)|-\pi(\xi))
e^{-\frac{1}{M_p}\varphi(\xi)}d\xi\biggr)\biggr\},\nonumber\\
\fl\alpha(\tau,\sigma)=\frac{1}{2} \biggl\{\ln\biggl(\frac{1}{2}
e^{\frac{1}{M_p}\varphi(\sigma-\tau)}+\frac{1}{2}
e^{\frac{1}{M_p}\varphi(\sigma+\tau)}+\frac{e^{-2\alpha_0}}{2M_p}\int_{\sigma-\tau}^{\sigma+\tau}(\pi(\xi)+|\pi(\xi)|)
e^{\frac{1}{M_p}\varphi(\xi)}d\xi\biggr)\nonumber\\\fl
+\ln\biggl(\frac{1}{2}
e^{-\frac{1}{M_p}\varphi(\sigma-\tau)}+\frac{1}{2}
e^{-\frac{1}{M_p}\varphi(\sigma+\tau)}+\frac{e^{-2\alpha_0}}{2M_p}\int_{\sigma-\tau}^{\sigma+\tau}(|\pi(\xi)|-\pi(\xi))
e^{-\frac{1}{M_p}\varphi(\xi)}d\xi\biggr)\biggr\}+\alpha_0,}
\label{resh}
\ee
where the following initial conditions  are taken
\be
\eqalign{ \phi(0,\sigma)=\varphi(\sigma),~~~
\alpha(0,\sigma)=\alpha_0=const, \\
\phi^\prime(0,\sigma)=e^{-2\alpha_0}\,\pi(\sigma),~~~~
\alpha^\prime(0,\sigma)=e^{-2\alpha_0}|\pi(\sigma)|.}
\label{cond}
\ee
Here $\varphi(\sigma)$ and $\pi(\sigma)$ are some functions
corresponding to the initial field and momentum, respectively.
 Under these initial conditions, the constraints $\mathcal H$ and
$\mathcal P$ are non-zero, but their relative magnitudes tend to zero with
$\alpha_0\rightarrow -\infty$:
\begin{eqnarray}
\frac{\mathcal
H}{e^{-2\alpha_0}\pi^2(\sigma)}=e^{4\alpha_0}\,\frac{(\partial_\sigma\varphi(\sigma))^2}{\pi^2(\sigma)}\rightarrow
0,\nonumber\\
\frac{\mathcal
P}{e^{-2\alpha_0}\pi^2(\sigma)}=e^{2\alpha_0}\,\frac{\partial_\sigma\varphi(\sigma)}{\pi(\sigma)}\rightarrow
0.
\label{asymt}
\end{eqnarray}
It should be noted, that the scale factor $a=\exp(\alpha)$  can be
collapsing (i.e. decreasing function of time) at some instant
but, certainly, remains always positive.

The quantization of the initial conditions consists in the
postulation of
 the commutation relation
\[
[\hat \pi(\sigma)\hat
\varphi(\sigma^\prime)]=-i\delta(\sigma-\sigma^\prime)
\]
for the operators on the right hand side  of Eq. (\ref{cond}). It
is worth to note, that $\hat \alpha(\tau,\sigma)$ is
\emph{c}-number at the initial moment of time.

Since the momentum constraint is disregarded, the
 WDW equation in the vicinity of $\alpha \rightarrow -\infty$ is written as
\begin{equation}
\left(\,\frac{\delta^2}{\delta
\alpha^2(\sigma)}-\frac{\delta^2}{\delta\varphi^2(\sigma)}\right)\Psi[\alpha,\varphi]=0,
\label{witt0}
\end{equation}
where the functional $\Psi[\alpha,\varphi]$  can be represented as
\begin{equation}
\Psi[\alpha,\varphi]=\int C[\pi]\,e^{\int\left(-i M_p|\pi(\sigma)|
\alpha(\sigma)+i\pi(\sigma)\varphi(\sigma)\right)d\sigma}\,\mathcal
D \pi,
\end{equation}
and a mean value of an operator becomes
\begin{equation}
\fl <\psi|\hat A[\alpha,\pi,i\frac{\delta}{\delta \pi}]|\psi>=\int
C^*[\pi]e^{i\alpha_0\int|\pi(\sigma)| d\sigma}\hat
A\,e^{-i\alpha_0\int|\pi(\sigma)|d\sigma}C[\pi]\,\mathcal D
\pi\biggl|_{\alpha_0\rightarrow -\infty }.
\end{equation}

Also, it is convenient to use the Wigner function \cite{groot}
\begin{equation}
\fl\wp[\pi,\varphi,\alpha_0]=\int C^*[2\pi-q]C[q]e^{
\int(-i\alpha_0|q(\sigma)|+i\alpha_0|2\pi(\sigma)-q(\sigma)|+2i(q(\sigma)-\pi(\sigma))\varphi(\sigma))d\sigma}{\mathcal
D}q,
\label{w}
\end{equation}
and the Weyl symbol $A[\pi,\varphi]={\mathcal W}[\hat A]$. The
latter can be calculated in accordance with the following rules:
\be
\eqalign{
 {\mathcal W}[\hat
\pi(\sigma)]=\pi(\sigma),~~~~ {\mathcal W}[\hat
\varphi(\sigma)]=\varphi(\sigma),\\
\fl W[\frac{1}{2}\left(\hat A\hat B+\hat B\hat
A\right)]=\cos\biggl(\frac{1}{2}\int\biggl(\frac{\delta}{\delta
\varphi_1(\sigma)}\frac{\delta}{\delta
\pi_2(\sigma)}\nonumber\\~~~~~~~~~~~~-\frac{\delta}{\delta
\varphi_2(\sigma)}\frac{\delta}{\delta
\pi_1(\sigma)}\biggr)d\sigma\biggr)A[\pi_1,\varphi_1]B[\pi_2,\varphi_2]\Bigl|_{\tiny
\begin{array}{c}
\pi_1(\sigma)=\pi_2(\sigma)=\pi(\sigma) \\
\varphi_1(\sigma)=\varphi_2(\sigma)=\varphi(\sigma) \end{array}}.}
\ee
Using the Weyl symbol and the Wigner function allows calculating a
mean value of an operator:
\begin{equation}
<|A|>=\int A[k,\varphi,\alpha_0]\wp[k,\varphi,\alpha_0]\,\mathcal
D k\,\mathcal D\varphi\bigl|_{\alpha_0\rightarrow -\infty}.
\label{meanw}
\end{equation}

To obtain some rough estimations, let's consider the classical
solution (\ref{resh}) of Eqs. (\ref{eqns1}) as the Weyl transform of
the solution of operator equations. In the next order, the quantum
corrections to the zero-order Weyl symbols will arise. The
investigation of the minisuperspace model \cite{prep2} has
demonstrated that the quantum corrections to the Weyl symbols are
not substantial and the quantum effects are contained mainly in the
Wigner function.

As the next step, it is required to interpret a sense of the
limit  $\alpha_0\rightarrow -\infty$. Let's consider the mean
value of $\hat \varphi(\sigma)=i\frac{\delta}{\delta
\pi(\sigma)}$, which can be written as
\be
\eqalign{ \fl <|\hat \varphi(\sigma)|>=\int C^*[\pi]e^{i
\alpha_0\int|\pi(\sigma)| d\sigma}\hat
\varphi(\sigma)\,e^{-i\alpha_0\int|\pi(\sigma)|d\sigma}C[\pi]\,\mathcal
D \pi\nonumber\\=\int C^*[\pi]\left(M_p\,\alpha_0
\frac{\pi(\sigma)}{|\pi(\sigma)|}+\hat
\varphi(\sigma)\right)C[\pi]\,\mathcal D \pi,}\label{56}
\ee
\noindent where $\alpha_0$ does not tend to $-\infty$ yet, since
the result is divergent at this stage. One can see from the Eq.
(\ref{56}) that, instead of using the state
$e^{-i\alpha_0\int|\pi(\sigma)|d\sigma}C[\pi]$ in calculation
of the mean value, one can take the $C[\pi]$ state but with the replacement
$\hat \varphi(\sigma)\rightarrow M_p\,\alpha_0
\frac{\pi(\sigma)}{|\pi(\sigma)|}+\hat \varphi(\sigma)$.
 As a rough approximation,
one can make this substitutions in the Weyl symbols
 (\ref{resh}) bearing in mind that the mean values of observabales are to be evaluated
 by the
ordinary quantum mechanical Wigner function (compare with
(\ref{w}))
\begin{equation}
\tilde \wp[\pi,\varphi]=\int C^*[2\pi-q]C[q]e^{
\int(2i(q(\sigma)-\pi(\sigma))\varphi(\sigma))d\sigma}{\mathcal
D}q.
\end{equation}

Taking into account that
\begin{eqnarray*}
\fl e^{\pm \frac{\varphi(\sigma)}{M_p}}\rightarrow e^{\pm
\left(\frac{\varphi(\sigma)}{M_p}+\alpha_0\frac{\pi}{|\pi|}\right)}
=e^{\pm
\frac{\varphi(\sigma)}{M_p}}\biggl(\frac{1}{2}e^{\pm\alpha_0}\biggl(
1+\frac{\pi}{|\pi|} \biggr)+ \frac{1}{2}e^{\mp\alpha_0}\biggl(
1-\frac{\pi}{|\pi|}\biggr)\biggr)
\end{eqnarray*}
 in the limit
$\alpha_0\rightarrow -\infty$, the Weyl symbols
 (\ref{resh}) are reduced to
\begin{eqnarray}
 \fl\tilde\phi(\tau,\sigma)=\frac{1}{2}M_p
\Biggl\{\ln\Biggl(
e^{\frac{1}{M_p}\varphi(\sigma-\tau)}\Biggl(1-\frac{\pi(\sigma-\tau)}{|\pi(\sigma-\tau)|}\Biggr)+
e^{\frac{1}{M_p}\varphi(\sigma+\tau)}\Biggl(1-\frac{\pi(\sigma+\tau)}{|\pi(\sigma+\tau)|}\Biggr)\nonumber\\
\fl~~~~
+\frac{2}{M_p}\int_{\sigma-\tau}^{\sigma+\tau}(\pi(\xi)+|\pi(\xi)|)
e^{\frac{1}{M_p}\varphi(\xi)}d\xi\Biggr)-\ln\Biggl(
e^{-\frac{1}{M_p}\varphi(\sigma-\tau)}\Biggl(1+\frac{\pi(\sigma-\tau)}{|\pi(\sigma-\tau)|}\Biggr)\label{resht0}\\+
e^{-\frac{1}{M_p}\varphi(\sigma+\tau)}\Biggl(1+\frac{\pi(\sigma+\tau)}{|\pi(\sigma+\tau)|}\Biggr)
+\frac{2}{M_p}\int_{\sigma-\tau}^{\sigma+\tau}(|\pi(\xi)|-\pi(\xi))
e^{-\frac{1}{M_p}\varphi(\xi)}d\xi\Biggr)\Biggr\}\nonumber,
\end{eqnarray}

\begin{eqnarray}
\fl \tilde \alpha(\tau,\sigma)=-\ln 4+\frac{1}{2}
\Biggl\{\ln\Biggl(
e^{\frac{1}{M_p}\varphi(\sigma-\tau)}\Biggl(1-\frac{\pi(\sigma-\tau)}{|\pi(\sigma-\tau)|}\Biggr)+
e^{\frac{1}{M_p}\varphi(\sigma+\tau)}\Biggl(1-\frac{\pi(\sigma+\tau)}{|\pi(\sigma+\tau)|}\Biggr)\nonumber\\
\fl~~~~+\frac{2}{M_p}\int_{\sigma-\tau}^{\sigma+\tau}(\pi(\xi)+|\pi(\xi)|)
e^{\frac{1}{M_p}\varphi(\xi)}d\xi\Biggr)+\ln\Biggl(
e^{-\frac{1}{M_p}\varphi(\sigma-\tau)}\Biggl(1+\frac{\pi(\sigma-\tau)}{|\pi(\sigma-\tau)|}\Biggr)\label{resht}\\+
e^{-\frac{1}{M_p}\varphi(\sigma+\tau)}\Biggl(1+\frac{\pi(\sigma+\tau)}{|\pi(\sigma+\tau)|}\Biggr)
+\frac{2}{M_p}\int_{\sigma-\tau}^{\sigma+\tau}(|\pi(\xi)|-\pi(\xi))
e^{-\frac{1}{M_p}\varphi(\xi)}d\xi\biggr)\Biggr\}.\nonumber
\end{eqnarray}

The ``reduced'' Weyl symbols  $\tilde\phi(\tau,\sigma)$,
$\tilde\alpha(\tau,\sigma)$ do not contain $\alpha_0$, that is all
divergencies arising under $\alpha_0\rightarrow -\infty$
remarkably cancel each other. This means that the normalization of
the wave function by choosing the $\alpha(\sigma)=\alpha_0\rightarrow
-\infty $ plane in the Klein-Gordon scalar product and
the quantization rules for the quasi-Heisenberg operators are
compatible.

Besides,  $\tilde\phi(\tau,\sigma)$, $\tilde\alpha(\tau,\sigma)$
satisfy the equation of motion (\ref{eqns1}). At the same time, Eqs.
(\ref{sv0}), (\ref{sv}) for the constraints do not obey the
``reduced'' solutions (\ref{resht0}),(\ref{resht}): \be \eqalign{
\fl e^{2\tilde \alpha}\left(-M_p^2\tilde\alpha^{\prime
2}-M_p^2(\partial_\sigma\tilde\alpha)^2+\tilde\phi^{\prime
2}+(\partial_\sigma\tilde\phi)^2\right)=\frac{1}{2}(\pi(\sigma
+\tau)\varphi^\prime(\sigma+\tau)\\~~~~~~~~~~~~~~~~~~~~~~~~~~~
~~~~~~~~~~~~~~~~~~~~~~~~~~~~~-\pi(\sigma-\tau)\varphi^\prime(\sigma-\tau)),}
\label{sv01} \ee \be \fl e^{2\tilde
\alpha}(M_p^2\tilde\alpha^\prime\partial_\sigma\tilde\alpha-\tilde\phi^\prime\partial_\sigma\tilde\phi)=
\frac{1}{2}\left(\pi(\sigma+\tau)\varphi^\prime(\sigma+\tau)+\pi(\sigma-\tau)\varphi^\prime(\sigma-\tau)\right).
\label{sv1} \ee More exactly, the ``Friedmann equation'' (\ref{sv0})
is satisfied at $\tau=0$, whereas Eq. (\ref{sv}) is never satisfied.
However, for the states $C[\pi]$ possessing an uniformity ``at the
mean'' such that the mean value of the function gradient
$\varphi^\prime(\xi) $ is zero, the constraints are satisfied ``at
the mean'', as well.

Unlike (\ref{resh}), the solutions (\ref{resht0}),(\ref{resht})
are discontinuous functions: there is a gap in the vicinity of
$\pi(\sigma)= 0$. Since it is difficult to deal with the piecewise
continuous functions, it is reasonable to restrict oneself to the
functionals $C[\pi]$ admitting only positive initial momentums.
The instance is
\[
\fl
C[\pi]=\left\{\begin{array}{c}\exp\left(\int\left(-u(\sigma-\sigma^\prime)
\pi(\sigma)\pi(\sigma^\prime)
-v(\sigma-\sigma^\prime)/(\pi(\sigma)\pi(\sigma^\prime))\right)d\sigma d\sigma^\prime\right), \pi(\sigma)>0\\
 0,~~~\pi(\sigma)<1,
\end{array}\right.
\]
where $u(\sigma)$, $v(\sigma)$ are some positive functions.
  For a positive  $\pi(\sigma)$, the
"reduced" Weyl symbols (\ref{resht0}), (\ref{resht}) result in
\begin{eqnarray}
\fl \tilde\phi(\tau,\sigma)=\frac{1}{2}M_p \biggl\{\ln\biggl(
\frac{2}{M_p}\int_{\sigma-\tau}^{\sigma+\tau}\pi(\xi)
e^{\frac{1}{M_p}\varphi(\xi)}d\xi\biggr)-\ln\left(
e^{-\frac{1}{M_p}\varphi(\sigma-\tau)}+
e^{-\frac{1}{M_p}\varphi(\sigma+\tau)}\right)\biggr\},\label{resht0n}\\
\fl \tilde \alpha(\tau,\sigma)=\frac{1}{2} \biggl\{\ln\biggl(
\frac{1}{2M_p}\int_{\sigma-\tau}^{\sigma+\tau}\pi(\xi)
e^{\frac{1}{M_p}\varphi(\xi)}d\xi\biggr)+\ln\left(
e^{-\frac{1}{M_p}\varphi(\sigma-\tau)}+
e^{-\frac{1}{M_p}\varphi(\sigma+\tau)}\right)\biggr\}.
\label{reshtn}
\end{eqnarray}

Now, one may calculate the evolution of observables. For
 $C[\pi]$, it is possible to take a squeezed state giving a small initial
 momentum $\pi(\sigma)$ and a large
 initial field $\varphi(\sigma)$ due to the uncertainty principle. However, the numerical calculations
 including the functional integration still turn out to be
 complicated.
Thus, one has to come to the next simplifications. Let's replace the
quantum  averaging by the spatial integration \be
<G>=\frac{\omega}{2 \pi}\int_0^{2\pi/\omega}
G(\pi(\sigma),\varphi(\sigma))~ d \sigma, \label{mm} \ee \noindent
and take the initial momentum and field in the form of
\begin{eqnarray}
\pi(\sigma)=p,\nonumber\\
 \varphi(\sigma)=A\, \cos(\omega \,\sigma),
 \label{in}
\end{eqnarray}
where $p>0$, $\omega\sim M_p$, and $A\ll M_p$ are some constants.

First of all, it is of interest to describe a vacuum state. However,
a scalar field is strongly ``mixed'' with a scale factor variable in
the system considered. Thus, one may hardly expect to find a state
like that of the ordinary QFT. It seems, that obtaining of such a
state needs to consider a number of scalar fields in the hope that
their mutual effect on a scale factor would result in an analog of
the QFT vacuum at a smooth classical background. But the analytical
solution does not  exist in the case of multiple scalar fields.
Thus, we  explore only a peculiar feature of vacuum state, namely, a
fluctuation with the frequencies up to the Planck mass.

Let us use (\ref{resht0n},\ref{reshtn},\ref{mm},\ref{in}) for
calculation of the mean values of the following quantities

\begin{eqnarray*}
<\alpha^\prime>\approx \frac{1}{2 \tau}+\dots\\
<\alpha^{\prime 2}>\approx \frac{1}{4 \tau^2}+\frac{A^2
\omega^2}{16 M_p^2}+\dots\\
<(\partial_\sigma \alpha)^{ 2}>\approx \frac{A^2 \omega^2}{16
M_p^2}+\dots\\
\frac{1}{M_p^2}<\varphi^{\prime 2}>\approx \frac{1}{4
\tau^2}+\frac{A^2 \omega^2}{16 M_p^2}+\dots\\
\frac{1}{M_p^2}<(\partial_\sigma \varphi)^{ 2}>\approx \frac{A^2
\omega^2}{16
M_p^2}+\dots,\\
\end{eqnarray*}
where ellipsis denotes the higher-order terms on $\frac{1}{M_p}$.
One can see, that $<\alpha^{\prime 2}>$ contains a fluctuating
term. On the other hand, the fluctuations do not affect an average
evolution described by $<\alpha^{\prime}>$. It should be noted,
that the sum
$\frac{1}{M_p^2}<\left(\phi^\prime(\tau,\sigma)\right)^2+\left(\partial_\sigma\phi(\tau,\sigma)\right)^2>$
equals approximately to
$<\left(\alpha^\prime(\tau,\sigma)\right)^2+\left(\partial_\sigma\alpha(\tau,\sigma)\right)^2>$.
Thus the fluctuating terms compensate mutually each other in the
mean value of the Hamiltonian constraint $\mathcal H$ and this
solves a part of the cosmological constant problem.

The idea of such a compensation was suggested in
\cite{wang,cong,bon0} but beyond a concrete quantization scheme
for the GR. In the sequel, the results of stochastic classical
analysis suggest to refuse this idea \cite{bon} and to assert that
the metric conformal fluctuations do not result in the
cancellation of vacuum energy. Fact of the matter is that it is
possible to choose the scale factor to be spatially uniform in the
classics owing to the diffeomorphism invariance. In other gauges,
the additional terms in the momentum-energy tensor appear. These
terms contain the scale factor derivatives, but they are the total
divergences and disappear after averaging in agreement with Ref.
\cite{isa}. However, the scale factor is a spatially nonuniform
operator in our quantization scheme and cannot become uniform
owing to a gauge transformation. Thus, one may conjecture that the
mutual compensation of fluctuating terms in GR can exist in the
quantum case in spite of its hiding in the classics.

 The second part of the cosmological constant problem is
to explain the accelerated universe expansion. The difference
$\frac{1}{M_p^2}<\left(\phi^\prime(\tau,\sigma)\right)^2-\left(\partial_\sigma\phi(\tau,\sigma)\right)^2>
\approx<\left(\alpha^\prime(\tau,\sigma)\right)^2-\left(\partial_\sigma\alpha(\tau,\sigma)\right)^2>$,
so that \be
<\alpha^{\prime\prime}>\approx\frac{2}{M_p^2}<\left(\partial_\sigma\phi(\tau,\sigma)\right)^2
-\left(\phi^\prime(\tau,\sigma)\right)^2>\approx-\frac{1}{2
\tau^2}.
\ee Hence one may conclude, that the acceleration of the universe
expansion is determined by the mean value of the difference of
potential and kinetic energies of field oscillators
\cite{conf,2,conf0}. However, there is no accelerated universe
expansion in the case considered here in contrast to Ref.
\cite{conf,2,conf0}, where the quantum fields against a classical
background are analyzed (also, see \cite{as} where the virial
theorem for a vacuum state is discussed).  As it has been mentioned,
a possible source of this problem is that a sole scalar field under
consideration is mixed too strongly with the background. Another
reason is that the estimation is too rough and does not take into
account an explicit structure of quantum states.

\section{Discission and Conclusion}

The quantization scheme for some particular case of the string in a
curved background space has been proposed. This scheme is adapted
for quantization of the systems possessing a global evolution. The
typical feature of such quantization is that we use the hyperplane
$a=0$ and the initial condition for the quasi-Heisenberg operators
at the initial moment of time, that allows a considerable
simplification of the WDW equation as well as the formulation of
operators. One may hope, that such a method can be useful for the
quantization of GR owing to solvability of the WDW equation in the
vicinity of $a=0$. It should be noted, that, if the wave function
obeys the WDW equation, there exists no an equivalent
Schr\"{o}dinger picture for this quantization scheme. Although
finding of a solution of operator equations for the quasi-Heisenberg
operators is extremely difficult task, it can be made numerically
hereafter.

Since the constraints are imposed as the operator equations at the
initial moment of time, a certain operator ordering is required in
the operator equations of motion to provide the conservation of
constraints during a quantum evolution. Otherwise, the constraints
will be violated during an evolution due to quantum noise arising
from the operator noncommutativity.

The heuristic estimations for a quantum string in the
(1+1)-dimensional background space with the weakened constraints
have been developed. It has been found, that the quantum
oscillations of the scale factor compensate the oscillations of
the scalar field. Thus, the cosmological constant problem
may not arise if the scale factor is quantized in a proper
way. It has been shown, that the mean value of the universe
acceleration expansion is proportional to the difference of
potential and kinetic energies of field oscillators, but this
topic needs more careful investigation hereafter.

\end{document}